%% file: main.tex
\documentclass[sigconf, nonacm]{acmart}
\usepackage{cleveref}

\makeatletter

\newcommand{\ACMemailhref}[2]{%
  \mbox{\href{#1}{#2}}%
}

\patchcmd{\@mkauthors@iii}
  {\href}
  {\ACMemailhref}
  {}{}

\patchcmd{\@mkauthors@iii}
  {\href}
  {\ACMemailhref}
  {}{}

\makeatother

\usepackage{amsmath}
\usepackage{mathtools}
\usepackage{amsthm}
\usepackage{enumitem}
\usepackage{bbm}
\usepackage{array}
\usepackage{booktabs}
\usepackage{dsfont}
\usepackage{subcaption}
\usepackage{comment}
\usepackage{xspace}
\usepackage{makecell}
\usepackage{graphicx}

\newlength{\contentheight}
\begin{document}
\include{defs}

\title{SAiFE-gym: Model-based Environments for Automated Market Making with Concentrated Liquidity}

\author{Georgios Chionas}
\affiliation{%
    \institution{University of Liverpool}
    \city{Liverpool}
    \country{UK}}
\authornote{Authors contributed equally to the paper.}
\email{g.chionas@liverpool.ac.uk}

\author{Charalampos Kleitsikas}
\affiliation{%
  \institution{King's College London}
  \city{London}
  \country{UK}}
\authornotemark[1]
\email{charalampos.kleitsikas@kcl.ac.uk}

\author{Stefanos Leonardos}
\affiliation{%
  \institution{King's College London}
  \city{London}
  \country{UK}
}
\email{stefanos.leonardos@kcl.ac.uk}

\author{Leandro S\'anchez-Betancourt}
\affiliation{%
\institution{University of Oxford}
 \city{Oxford}
 \country{UK}}
\email{leandro.sanchezbetancourt@maths.ox.ac.uk}

\author{Carmine Ventre}
\affiliation{%
   \institution{King's College London}
  \city{London}
  \country{UK}
}
\email{carmine.ventre@kcl.ac.uk}

\renewcommand{\shortauthors}{Chionas et al.}

\begin{abstract}
We present \saife, a Python module that provides a collection of simulation environments for studying trading problems in constant product markets (CPMs) with concentrated liquidity (CL).
These markets give liquidity providers (LPs) granular control over how their capital is allocated and enable them to adjust their range of liquidity provision dynamically based on market conditions, which in turn determines how they earn fees. 
We decompose the microstructure of \clmms into interactive components that allow researchers and practitioners to capture various economic settings. 
We employ a vectorized approach to optimize our environments, making them scalable for high-dimensional reinforcement learning (RL) workflows, suitable for learning dynamic liquidity provision strategies.
We demonstrate the benefits of our environments by evaluating the performance of RL agents in \clmms under various market conditions. 

\end{abstract}

\maketitle

\section{Introduction}
Automated market makers (AMMs) are some of the most used applications for trading built on  blockchains. These markets were proposed as an alternative to limit order books (LOBs), tailored to the computational and storage limitations of blockchains. AMMs, which mainly operate as constant function markets (CFMs), dictate the rules of clearing demand and supply through bonding curves, with the most dominant example being the constant product bonding curve. 
Despite their simple design and the inefficiencies in their first iterations, AMMs have facilitated \$4 trillion in trading volume across Ethereum and other Ethereum-compatible networks \cite{uniswap2026volume}.

Consequently, AMMs have attracted the interest of researchers within the mathematical finance community \cite{Angeris20CFMMs, Cartea2024defi, Cartea25Execution}. Most of these works study stylized models, either from the liquidity provider (LP) perspective \cite{Cartea2024defi} or from liquidity takers (LTs) \cite{Cartea25Execution, Angeris22}, by capturing the AMM mechanics associated with their model specific assumptions. Despite the mathematical elegance of those methods, they are often low-dimensional to accommodate analytical solutions and, thus, are unable to capture the high-dimensionality observed in practice. Reinforcement learning (RL), in turn, has emerged as a paradigm for tackling such high-dimensional control problems. 
Nevertheless, we observe that there is fragmentation in model-based simulation environments tailored to the AMM microstructure, and to this day, there is not an open-source, unified and extensible collection of environments available to the research community.

We introduce \saife,
a collection of RL environments, that assembles components which govern each AMM case study. 
Environments have the property that once the skeleton is in place, the research community and practitioners can parallelize across different domains. For that reason, RL environments sit at the centre of current AI progress, where each new one expands what we can train, study, and evaluate.
By lowering the barriers to conducting research on AMMs, our \saife contributes directly to the intersection between crypto and artificial intelligence \cite{allen2026crypto}. 

The goal of this work is to assemble different aspects of the AMM microstructure in an extensible way while also demonstrating the potential of RL for solving this type of stochastic control problem. Broadly, the AMM microstructure is built from several interacting components, namely the evolution of market prices, the order flow and its toxicity, the price impact and the liquidity distribution, and the frictions imposed by blockchains, namely the gas costs for interacting with them and their inherent latency. 
Our main architecture design for the RL environments is a vectorized approach, whereby a batch of trajectories runs in parallel, aggregating the results. This design choice significantly improves computational efficiency (wall-clock efficiency) while also providing more stable estimates for policy optimization. The summary of our contributions is given below.

\paragraph{Our Contributions.}
\begin{enumerate}[leftmargin=*, nosep]
    \item We provide a unified set of RL environments that replicate the mechanics of AMMs under a wide range  of market dynamics. This is a key strength of our approach as it supports picking and choosing different components of modelling that  simulate various economic settings. 
    \item We  connect our environments to Stable Baselines 3 (SB3) \cite{stablebaselines3}, which provides a rich suite of state-of-the-art RL algorithms. These algorithms allow a ``plug-and-play'' approach to train RL agents on previously unstudied problems, while also enabling users to develop their own customizable agents. 
    \item Following the design principles of \cite{Jerome23mbt-gym}, we develop a high degree of parallelism in our collection of RL environments, which work on top of common existing multiprocessing methods, achieving a level of efficiency for training RL agents that makes the learning tasks feasible even for conventional machines.
    \item To demonstrate the applicability of our environments, we evaluate the performance of RL agents under various market conditions. This approach, known as \textit{domain randomization}, has found great applicability in simulation-to-real work, see \cite{Peng18sim}.
\end{enumerate}
The module is available on 
\href{https://github.com/giorgoschionas/SAiFE_gym}{GitHub}.

\section{Related Work}
\paragraph{Automated Market Makers.}
The computational efficiency of using invariant functions (fixed trading functions)  together with the computational constraints of the early blockchains led simple AMMs, in the form of constant product markets, like Uniswap v2 \cite{Adams20v2}, to become a cornerstone in decentralised finance. 
The work in \cite{Angeris20CFMMs} generalized the notion of the constant product invariant to the broader class of constant function market makers (CFMMs) and characterised conditions under which arbitrage aligns the price quoted by a CFMM with the (true) price of an asset on an external market, thus establishing the usefulness of CFMMs as price oracles. 

Despite their widespread adoption, simple AMMs such as Uniswap v2 suffer from capital inefficiency, since a substantial part of the provided liquidity is never used. Newer protocols such as Uniswap v3 \cite{Adams21v3} enable LPs to specify price ranges within which their liquidity is active. This mechanism, known as concentrated liquidity (CL), gives LPs more granular control over the allocation of their capital. Consequently, Uniswap v3 significantly expands the LPs' action space, which was previously considered passive. 

This transition has sparked a significant line of research on profitable strategies for LPs in \clmms. We outline notable works that inspired the collection of environments presented in this work. 
The authors in \cite{Fan23StrategicLP, Fan22differential} study dynamic liquidity provision strategies in Uniswap v3 and provide insights on how aspects of a Uniswap v3 contract, notably the partition of price space, have implications for LPs' profit.
The authors in \cite{Cartea2023predictable} decompose the LP's losses in \clmms into two parts; the convexity cost, arising from the convexity of the pool's invariant function, and the opportunity cost, arising from the LP's locking assets in the pool instead of investing them in a risk-free account. Their subsequent work in \cite{Cartea2024defi} formulates dynamic liquidity provision as a stochastic control problem and derives a closed-form solution, under which the LP reallocates its accumulated wealth continuously across intervals as the price evolves. However, their model abstracts away from frictions inherent to blockchain, mainly the gas fees and latency. 

\paragraph{Reinforcement Learning for Market Making.} In recent years, RL has gained considerable traction in tackling model-based problems in high-frequency trading. This is because market making in LOBs, as well as dynamic liquidity provision in \clmms, can naturally be formulated as stochastic control problems. RL offers a complementary method to numerical solutions, which usually suffer from the curse of dimensionality and therefore struggle to capture realistic models. While there is substantial literature and open-source infrastructure for model-based RL environments in LOB trading, similar tools for the newer exchange mechanisms of AMMs remain limited. 
   
In traditional finance, the work in \cite{Spooner18MM} introduces one of the earliest RL frameworks for market making in a LOB simulator and shows that inventory-aware RL agents can outperform simple known benchmarks. 
Closely related in spirit to our work, the work in \cite{Jerome23mbt-gym} provides a modular suite of Gym-compatible environments for model-based LOB trading that supports picking different components of existing models in the mathematical finance literature. 
However, in that work, a reduced-form model of the LOB is implemented by abstracting away from the statefulness of the entire order book. In contrast, our implementation of the \clmm  mirrors the respective pools that are implemented in practice.

Agent-based simulators provide an alternative to model-based simulators. Rather than specifying the market dynamics directly, agent-based simulators generate these dynamics through the interaction of agents. 
Some notable works that implement agent-based simulators to train RL agents in financial markets include \cite{kumar2020deep}, \cite{amrouni2021abides}, and \cite{Mascioli24RLagent-based}.
These are some of the most important works that demonstrate the effectiveness of RL in traditional finance settings. Our work is one of the first to provide a modular infrastructure for testing RL agents in AMMs.  

\section{Design of the Module}
\label{sec:design-module}
In the following sections, we give a detailed description of the implemented RL environments, including the mechanics of a \clmm and the microstructural dynamics that specify the environments. 
Formally, the interaction of the LP with the \clmm can be viewed as a stochastic control problem over a continuous time horizon $\mathcal{I} = [0,T]$, with terminal time $T$. In our RL environments, we approximate this problem on a discrete time grid with $n$ steps. At each time step, every RL environment performs the observation phase, where the LP agent observes market information, action phase, where the agent takes a new action, market operation phase, where the market processes the new orders and time advancement phase. These phases are implemented in \textit{AMMEnvironment}, from which all RL environments inherit. 
We then present experiments demonstrating the capabilities of our collection of environments. Finally, we discuss the implications of our work and outline directions for future research.

\subsection{Constant product markets and concentrated liquidity}
\paragraph{Constant Product Markets.}
Let $X$ denote a numéraire asset, such as USDC, and let $Y$ denote
a risky asset, such as ETH. A liquidity pool for the pair
$(X,Y)$ quotes a marginal exchange rate $Z$, measured in units of
asset $X$ per unit of asset $Y$. Let $q^X,q^Y>0$ denote the pool reserves of assets $X$ and $Y$,
respectively. The reserves of a CPM satisfy
the invariant
\begin{equation}
\label{eq:cpm-invariant}
  f(q^X,q^Y)
    =
    q^Xq^Y
    =
    \kappa^2,
\end{equation}
where $\kappa>0$ is the liquidity parameter, or depth, of the pool.
In the absence of fees, LT trades move the reserves along the level set
defined in \eqref{eq:cpm-invariant}. Differentiating the invariant
shows that the marginal exchange rate (the spot price) is $ Z= -\frac{\mathrm{d}q^X}{\mathrm{d}q^Y}=\frac{q^X}{q^Y}$.
Thus, $Z$ is the pre-fee marginal price of asset $Y$ in units of
asset $X$ for an infinitesimal trade. The CPM charges a proportional trading fee $\fee\in[0,1)$ on the input amount. Consequently, only the fraction $1-\fee$ of a trade's gross input is used in the swap calculation.

Consider now an LT who submits $\Delta x>0$ units of asset $X$ and
receives $\Delta y^{\mathrm{out}}>0$ units of asset $Y$. The output
amount is determined by the invariant equation
$
\left(q^X+(1-\fee)\Delta x\right)
\left(q^Y-\Delta y^{\mathrm{out}}\right)
=
q^Xq^Y.
$
Solving for $\Delta y^{\mathrm{out}}$ gives
\begin{equation*}
\label{eq:cpm-x-to-y}
    \Delta y^{\mathrm{out}}
    =
    q^Y
    -
    \frac{q^Xq^Y}
    {q^X+(1-\fee)\Delta x}.
\end{equation*}

Conversely, suppose that the LT submits $\Delta y>0$ units of asset
$Y$ and receives $\Delta x^{\mathrm{out}}>0$ units of asset $X$.
The invariant equation is
$\left(q^X-\Delta x^{\mathrm{out}}\right)
\left(q^Y+(1-\fee)\Delta y\right)
=
q^Xq^Y,$ which implies that
\begin{equation*}
\label{eq:cpm-y-to-x}
    \Delta x^{\mathrm{out}}
    =
    q^X
    -
    \frac{q^Xq^Y}
    {q^Y+(1-\fee)\Delta y}.
\end{equation*}

\paragraph{Concentrated Liquidity.}
In a \clmm, an LP can restrict a liquidity position to a specified
interval of marginal exchange rates $(Z^\ell,Z^u]$. The endpoints of
the position are selected from a finite, ordered grid
\begin{equation}
\label{eq:grid}
    \mathcal{G}
    =
    \big\{
        Z^{(-N)}=\underline Z,\,
        Z^{(-N+1)},\ldots,
        Z^{(0)},\ldots,
        Z^{(N-1)},\,
        Z^{(N)}=\overline Z
    \big\}, 
\end{equation}
for some $N\geq 1$. Each pair of neighbouring grid points  $\left(Z^{(i)},Z^{(i+1)}\right]$ is called a \emph{tick range}. Within a tick range, the pool satisfies a translated constant-product
invariant. Equivalently, the \clmm is a segmented CPM tailored to the liquidity in each tick range.

When opening a position, the LP selects a range of rates $(Z^\ell,Z^u]$ and
a liquidity amount $\tilde{\kappa}>0$. These variables, together with the current marginal
exchange rate $Z$, determine the corresponding quantities $x$ and $y$ of assets $X$ and $Y$ respectively, according to
\begin{equation}
\small
\label{eq:uni-v3-positions}
(x,y)=
\begin{cases}
\left(
0,\,
\tilde{\kappa}
\left(
    (Z^\ell)^{-1/2}-(Z^u)^{-1/2}
\right)
\right),
& Z\leq Z^\ell, \\[0.6em]

\left(
\tilde{\kappa}
\left(
    Z^{1/2}-(Z^\ell)^{1/2}
\right),\,
\tilde{\kappa}
\left(
    Z^{-1/2}-(Z^u)^{-1/2}
\right)
\right),
& Z^\ell<Z\leq Z^u, \\[0.6em]

\left(
\tilde{\kappa}
\left(
    (Z^u)^{1/2}-(Z^\ell)^{1/2}
\right),\,
0
\right),
& Z>Z^u.
\end{cases}
\end{equation}
In words, when $Z\leq Z^\ell$, the position consists only of asset $Y$, while when $Z>Z^u$, it consists only of asset $X$. When $Z^\ell<Z\leq Z^u$, the position is active
and contains quantities of both assets.

In a tick range $i$, the total liquidity depth $\ka_i$ is obtained by summing the liquidity amounts $\tka_i$ of each individual position in tick range $i$. Thus, the whole liquidity depth profile is captured by 
\begin{equation}
\label{eq:total-liquidity}
    \boldsymbol{\kappa}
    =
    \left(\kappa_i\right)_{i=-N}^{N-1}.    
\end{equation}
The formulas in \eqref{eq:uni-v3-positions} and \eqref{eq:total-liquidity} give a reparameterization of the state of the CPM with CL from the quantities $x,y$ of assets $X,Y$ respectively, to the total liquidity depth in each tick range and the marginal exchange rate. The latter is more convenient, since, an LP action changes only the liquidity depth profile \eqref{eq:total-liquidity}, while an LT action changes only the marginal exchange rate $Z$.

Lastly, the trading fees are stored in each tick range, and are distributed pro-rata to each LP. For example, if the LP's position with depth $\tka$ is in a tick range with total depth $\kappa$, then for every liquidity taking trade that pays in fees an amount of $p$, the LP  earns the amount of $\tilde{p} =  \frac{\tilde{\ka}}{\kappa} p \,  \mathbf{1}_{Z^{\ell} < Z \le Z^u}$.

\subsection{Mid-price processes}
\label{subsec: midprice}
In our RL environments, we follow the standard assumption in the literature on AMMs, under which the price formation of the risky asset $Y$ is exogenous to the AMM.
Concurrently, the AMM quotes its own price for the risky asset $Y$, the marginal exchange rate $Z$.  
\paragraph{External Midprice (Exogenous)}
We call this process the external mid-price and denote it by $S_t$. There are various mid-price models implemented in our environments, such as the geometric Brownian motion which is the solution of the stochastic differential equation 
\begin{equation} 
\label{eq:GBM}
    dS_t = \mu S_t dt + \sigma S_t dW_t,\qquad S_0>0,
\end{equation} 
with drift $\mu$, volatility $\sigma>0$, and where $W_t$ is a Brownian motion. 
Another implemented model is the arithmetic Brownian motion, where the mid-price can be written as
\begin{equation*}
S_t = S_0 + \mu t+ \sigma W_t, \qquad S_0>0.
\end{equation*}
We also provide an arithmetic midprice model with an Ornstein–Uhlenbeck alpha signal, from the algorithmic trading literature.
\begin{align*}
    dS_t &= \alpha_t dt + \sigma dW_t,\qquad S_0>0,\\
    d\alpha_t &= -k^\alpha \alpha_t dt + \sigma^\alpha dW^\alpha_t,\qquad \alpha_0\in\mathbb{R},
\end{align*}
where $k^\alpha$ is the mean-reversion rate, $\sigma^\alpha>0$ is the volatility of the signal, and $W^\alpha$ is a Brownian motion independent of $W$.
\paragraph{Marginal Exchange Rate (Endogenous)}
The marginal exchange rate $Z$ is modelled as a pure jump process. We assume that the marginal exchange rate takes values from the same discrete grid \eqref{eq:grid}. 
The evolution of the marginal exchange rate is endogenous because it depends on the state of the environment as explained next.

\subsection{Arrival processes}
\label{subsec:arrivals}
Currently in AMMs, the formation of the marginal exchange rate is mainly due to LTs' activity. 
In our RL environments, we decouple the LTs' activity into an exogenous and an endogenous component. 
\paragraph{Noise Traders, Liquidity-attracted Traders and Informed Traders}
The entire LT action is captured by the next model for intensities. 
\begin{equation}
\lambda_t^{\pm} = \max\left\{\alpha_0,\, \alpha_1 + \alpha_2 K^{\pm}_{t-} \pm \alpha_3\left(Z_t-{S_{t-}}\right)\right\}
\label{eq:linear_intensities}
\end{equation}
where $\lambda^{+}_t$ (resp. $\lambda^{-}_t$) is the intensity for LT sells (resp. LT buys). The linear model \eqref{eq:linear_intensities}, which was introduced in \cite{Aqsha2025equilibrium}, captures three stylized facts. The parameter $\alpha_1$ is the baseline intensity of order arrivals, which corresponds to noise traders, and this part is exogenous to the state of the AMM.
The parameter $\alpha_2$ models the relationship between arrivals and depth (which means that higher depth implies more arrivals) and $\alpha_3$ captures the impact of LT action by informed traders who align the marginal exchange rate $Z$ with the external mid-price $S$.\footnote{Linear intensity models are widely used  in the literature, see e.g., \cite{Stoikov2009option}. Writing the dynamics of the AMMs à la Avellaneda-Stoikov was first proposed in \cite{Cartea23AMMDesigns}.} Lastly, $\alpha_0>0$ is a safeguard parameter to prevent intensities from becoming negative.
More specifically, when  $S>Z$ (resp. $S<Z$), all else being equal, we expect a higher (resp. lower) buying intensity and a lower (resp. higher)
selling intensity, given that the marginal exchange rate is underpriced (resp. overpriced). The subscript $t-$ in \eqref{eq:linear_intensities} represents the time just before $t$ given that we are working with processes that jump.

\begin{figure}[t]
    \centering
    \includegraphics[width=\columnwidth]{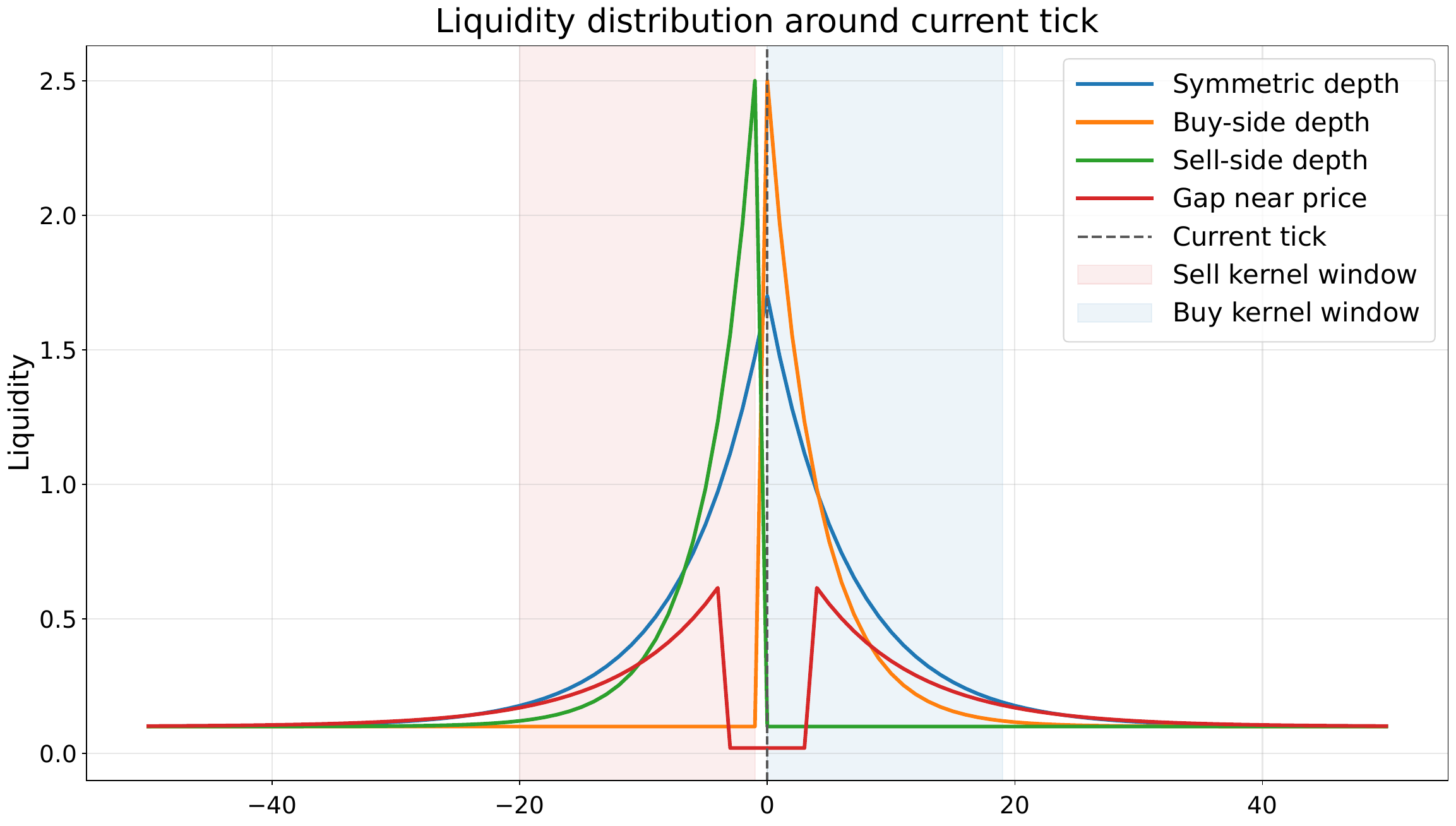}
    \caption{Smoothed kernels with exponential weights capturing the effect of the neighbouring liquidity of the current marginal exchange rate to the arrivals intensity.}
    \label{fig:liquidity-kernels}
\end{figure}

The liquidity statistic $K^{\pm}_{t-}$ in \eqref{eq:linear_intensities} is implemented through exponential kernels $\mathcal{K}^{\pm}$, which are illustrated in \Cref{fig:liquidity-kernels}. 
These kernels encode how the liquidity standing ahead of the current tick $i$ affects future arrivals. In particular, we implement
\begin{equation}
\label{eq:liquidity-kernels}
    \mathcal{K}^+_i = \sum\limits_{h=0}^{H-1}w_{h+1} \ka_{i+h}  \quad \text{and } \quad \mathcal{K}^-_i = \sum\limits_{h=0}^{H-1} w_{h+1} {\ka_{i-h}}
\end{equation}
where the weights $w$ determine how liquidity in ticks away from the current tick $i$ affects future arrivals. 
Through this channel, the LP's action influences the future order flow.\footnote{These liquidity kernels are similar in spirit to the queue-reactive model studied in LOBs \cite{Huang02012015}, but, to the best of our knowledge, this is the first time considered in the context of \clmms.} 

In our RL environments, the parameters $\alpha_1,\alpha_2,\alpha_3$ and the kernels $\mathcal{K}^{\pm}$ can be modulated to represent different AMM economic settings, or they may be estimated from empirical data, as  in \cite{Aqsha2025equilibrium}.

\subsection{Price Impact}
\label{subsec:price-impact}
The price impact refers to the function that connects the marginal exchange rate before and after the execution of an LT trade. We provide two classes. Recall that the marginal exchange rate $Z$ is implemented as a jump process and it takes values on the grid \eqref{eq:grid}. 
\paragraph{Depth-independent price impact.} 
This is a reduced-form model whereby each LT trade pushes the marginal exchange rate to the neighbouring tick based on the direction of the trade. If the marginal exchange rate is at tick $i$, a buy LT trade (resp. a sell trade) pushes $Z$ to tick $i+1$ (resp. to tick $i-1$). This reduced-form price impact model in \clmms has been used in the works \cite{Fan22differential, Fan23StrategicLP}.

\begin{table*}[t]
\centering
\scriptsize
\setlength{\tabcolsep}{4pt}
\renewcommand{\arraystretch}{1.15}

\begin{tabular}{
    p{0.18\textwidth}
    p{0.37\textwidth}
    p{0.37\textwidth}
}
\hline
\textbf{Economic component} &
\textbf{Role in the environment} &
\textbf{Implemented models} \\
\hline

External midprice &
Generates the exogenous market reference price $S_t$ observed by the AMM and the LP. &
\makecell[l]{
    \texttt{BrownianMotionMidpriceModel}; \\
    \texttt{GeometricBrownianMotionMidpriceModel}; \\
    \texttt{OrnsteinUhlenbeckMidpriceModel}
} \\
\hline

Order arrivals &
Generates LT action and updates its intensity from AMM state. &
\makecell[l]{
    \texttt{LiquidityKernelArrivalModel}
} \\
\hline

Price impact &
Maps LT trades into jumps of the marginal exchange rate. &
\makecell[l]{
    \texttt{OneTickUniswapV3PriceImpact}; \\
    \texttt{LiquidityDepthUniswapV3PriceImpact}
} \\
\hline

Reward function &
Defines the objective optimized by the agent. &
\makecell[l]{
    \texttt{PnL}; \\
    \texttt{RunningInventoryPenalty}
} \\
\hline

\end{tabular}

\caption{
Components in \texttt{SAiFE\_gym}. Each setting is exposed through a
small interface, enabling plug-and-play of different components while
keeping the remaining AMM engine fixed.
}
\label{tab:saife-modular-components}

\end{table*}

\paragraph{Depth-dependent price impact.}
In practice, price impact depends on the curvature of the CPM and the concentration of liquidity around the current marginal exchange rate. In order to capture these factors, we compute a smoothed 
forward and backward depth of the \clmm, over a given horizon of neighbouring ticks. 
The smoothing approach is similar in spirit to the kernel implementation for the order arrivals \eqref{eq:liquidity-kernels}. We choose a smoothing horizon $H$ and non-negative weights
$w_1, w_2, \dots , w_H$ and $\sum\limits_{j=1}^Hw_j=1$.
The smoothed forward depth (resp. smoothed backward depth) at tick $i$ are
\begin{equation}
\label{eq:depth-forward-backward}
\begin{aligned}
    &D^+_i = \sum\limits_{h=0}^{H-1}w_{h+1}\ka_{i+h} \left( (Z_{i+h+1})^{1/2}-(Z_{i+h})^{1/2}\right), \\
    &D^-_i = \sum\limits_{h=0}^{H-1}w_{h+1} \ka_{i-h-1}\left((Z_{i-h-1})^{-1/2}- (Z_{i-h})^{-1/2}\right)
\end{aligned}
\end{equation}
where $\ka_j$ is the total liquidity at tick $j$. A buy trade (resp. a sell trade) causes a price impact tailored to the smoothed forward depth (resp. smoothed backward depth). In particular, the size $Q$ of the LT trade (in units of the numeraire asset $X$) is drawn from a custom distribution (e.g. a lognormal distribution).\footnote{For a sell LT trade, the size in units of asset $Y$ is given by dividing the size $Q$ by the current external price $S$.} The price impact is the number of jumps of the marginal exchange rate in the grid \eqref{eq:grid} and it is calculated by $m = \lfloor Q/D_i^{\pm} \rfloor$. Therefore, when the marginal exchange rate is at tick $i$, a buy (resp. sell) trade of size $Q$ moves it to tick $i +m$ (resp. to tick $i - m$).

\subsection{Blockchain Frictions}
\paragraph{Gas Costs}
Our RL environments capture the cost of rebalancing as a fixed transaction cost. The fixed transaction cost corresponds to the \textit{gas fees} paid by the LP to the network for interacting with the blockchain. Gas fees are mostly a flat fee paid to the blockchain and do not depend on LP's specific position. Thus, gas fees scale with the frequency of rebalancing. In particular, one can write the cumulative cost up to time $t$ as
\begin{equation}
\label{eq:impulse-lp-cost}
    C_t=
    \sum\limits_{\imp\le t}
    g_r, \notag
\end{equation}
where $r$ is the number of times at which a rebalancing took place and $g_r>0$ is the gas cost associated with the rebalancing.
\paragraph{Latency}
In blockchain environments, there is inherent latency, which prevents the LP from reallocating her liquidity and rebalancing her assets continuously. In our RL environments, we capture latency by specifying the number of decisions throughout training and the evaluation time. In other words, we allow deterministic latency. 

Overall, all the components described in this section are summarized in \Cref{tab:saife-modular-components}.

\subsection{Agents}
We have implemented custom agents by primarily focusing on liquidity provision in \clmms. All the agents inherit from a base class \texttt{Agent} with the main function being \texttt{get\_action(state)}. 
\paragraph{Liquidity Providers}
We implement two ways in which the LP agent can interact with the \clmm that aligns with the admissible actions of an LP in \clmms such as Uniswap v3 and v4. 
Firstly, the agent can open an active position, where she decides the lower and upper tick $(Z^{\ell}, Z^u]$ such that the current marginal exchange rate $Z \in (Z^{\ell}, Z^u]$.
Secondly, the LP agent can approximate a limit order by opening a position $(Z^{\ell}, Z^u]$ outside of the current marginal exchange rate, whereby $Z\le Z^{\ell}$ or $Z> Z^u$. This is essentially a single-sided asset provision and in particular, provision of the risky asset $Y$ in the former case and provision of the numeraire asset $X$ in the latter case as defined in \eqref{eq:uni-v3-positions}. Whenever the LP chooses to rebalance, she first withdraws her holdings in assets $X,Y$, collects the accumulated fees from her last position and then reinvests her entire wealth in the new position.

\paragraph{Liquidity Takers}
Although our main focus on the presented RL environments is liquidity provision, we also provide an LT agent for statistical arbitrage against the \clmm.\footnote{We refer to the work \cite{Cartea25Execution} for a detailed treatment on liquidity taking in CPMs.} The control of the LT agent is the signed trading speed $v$ at which the agent is buying or selling the asset $Y$. In particular, if $v>0$ (resp. $v<0$), the agent sells (resp. buys) the risky asset $Y$. If $v=0$, the LT agent does not trade.
For each timestep $dt$, the trading speed $v$ implies the size $q$ of the LT's trade is $v \cdot dt$. The LT's execution cost is specified through the price impact that depends on the LT trade size $q$, which is calculated by the \textit{Depth-dependent price impact} given in \Cref{subsec:price-impact}.

\subsection{Reward Functions}
Let $T>0$ be the terminal time of the simulation and $\mathcal{I} =[0, T]$. 
The reward PnL (the risk-neutral reward) consists of the mark-to-market value of the LP's position. Let $\left(x_t \right)_{t \in \mathcal{I}}$ be the process of the agent's holdings in the numeraire asset, $\left(y_t \right)_{t \in \mathcal{I}}$ be the process of the agent's holdings in the risky asset and $\left(S_t \right)_{t \in \mathcal{I}}$ be the external midprice process. The final mark-to-market agent's value is $V_T = x_T +y_T S_T$. For the LP agent, the gas fees are subtracted from $x_t$, at every rebalancing step. The PnL is the change in the agent's value between steps and therefore it can be written as $V_T - V_0$.
\paragraph{Risk Aversion}
We implement a reward that captures risk aversion through a penalty of holding the risky asset $Y$ (an \textit{inventory} penalty). We use the Cartea-Jaimungal criterion \cite{Cartea17Uncetrainty}.
The reward \textit{RunningInventoryPenalty} is calculated as 
\begin{equation} 
\label{eq:running_inv}
 V_T- V_0 - \eta \left(y_T-\hat{y} \right)^2
     - \pen \int_0^T(y_t - \hat{y})^2dt,
\end{equation}
where the constants $\eta, \pen \in \mathbb{R}^+$ determine the terminal and running penalty respectively from deviating from some target level $\hat{y}$. The units of $\eta, \pen$ are such that the penalty terms are in units of $X$, while the parameter $\pen$ quantifies the urgency of the agent to liquidate the risky asset $Y$. 

\section{Reinforcement Learning}
In this Section, we describe how the AMM engine, presented in \Cref{sec:design-module}, works along with RL workflows. This is achieved through the wrappers that we outline in \Cref{subsec:wrappers}. Subsequently, in \Cref{subsec:vectorized-approach}, we present the custom vectorized approach that we follow across all the environments, which enables noticeably faster convergence in the agents' rewards. Lastly, in \Cref{subsec:domain-rnd}, we demonstrate a use case of our collection of environments in domain randomized reinforcement learning; specifically, the evaluation of RL agents when training is driven by uncertainty in the market parameters embedded in the stochastic processes presented in \Cref{sec:design-module}.

\begin{figure}[h]
    \centering
    \includegraphics[width=\linewidth]{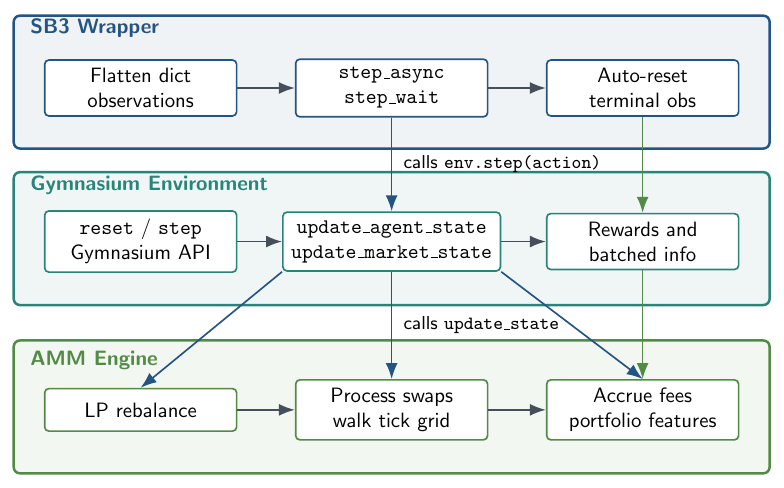}
    \caption{Deep RL wrapper workflow}
    \label{fig:saife-architecture}
\end{figure}

\subsection{Wrappers}
\label{subsec:wrappers}
\saife provides Gymnasium APIs for single agent deep reinforcement learning in the simulated \clmm. \Cref{fig:saife-architecture} illustrates how a learning agent interacts with the AMM engine through the SB3 wrapper. 
From the learning agent's perspective, the AMM mechanics, the external midprice, and the arrivals are all part of the environment. At each decision time, the learning agent observes the current state, chooses an action, and passes this action to the environment through the \texttt{step} interface. 
The environment then updates the AMM state by applying the agent's action together with the simulated market dynamics. The resulting transition follows the standard Gymnasium convention; the \texttt{step(action)} returns the next observation, reward, and episode termination information. In this way, the wrapper layer exposes transition tuples consisting of observation, action, and next observation, while remaining agnostic to the specific RL algorithm used. This wrapper is the main compatibility layer between the AMM engine and the SB3 training algorithms. Next, we briefly describe the functionality of the main agent-specific wrappers that are implemented.

\paragraph{StructuredMultiDiscreteVecEnv} This wrapper re-parameterizes the action space from \texttt{[lower\_offset, upper\_offset, hold]}, which is given in \eqref{eq:uni-v3-positions}, to \texttt{[center, half\_width, hold]}. The former parameterization is Uniswap v3-native, while the latter one is more suitable for RL training since the decisions are separable.

\paragraph{DecisionStrideEnv}
This wrapper separates the market decision frequency from the agent's decision frequency. The role of this wrapper is to capture the latency inherent in blockchain environments. Intuitively, it transforms a high-frequency control problem into a lower-frequency decision problem.\footnote{Dynamic liquidity provision strategies are often slower than the underlying market simulation. The AMM pool might update every time step, but an LP would not necessarily rebalance after every LT trade.}
Through this wrapper, one can customize the number of decisions per episode the agent will take in equally split time intervals. For example, by setting \texttt{K\_stride}=0, we recover the continuous-time setting where the market decision frequency aligns with the agent's decision frequency.

\paragraph{ReduceStateSizeWrapper}
The implementation of the \clmm is stateful enough; hence, one could customize the observation space of the agent. We do offer several standard features that can be used as components of the observations, namely:
\begin{enumerate}[label=\roman*]
    \item time remaining $T-t$,
    \item the mispricing $S_t-Z_t$ between the external midprice and the marginal exchange rate,
    \item the LP's current position,
    \item the accrued fees.
\end{enumerate}

\begin{figure}[t]
    \centering
    \includegraphics[width=0.8\columnwidth]{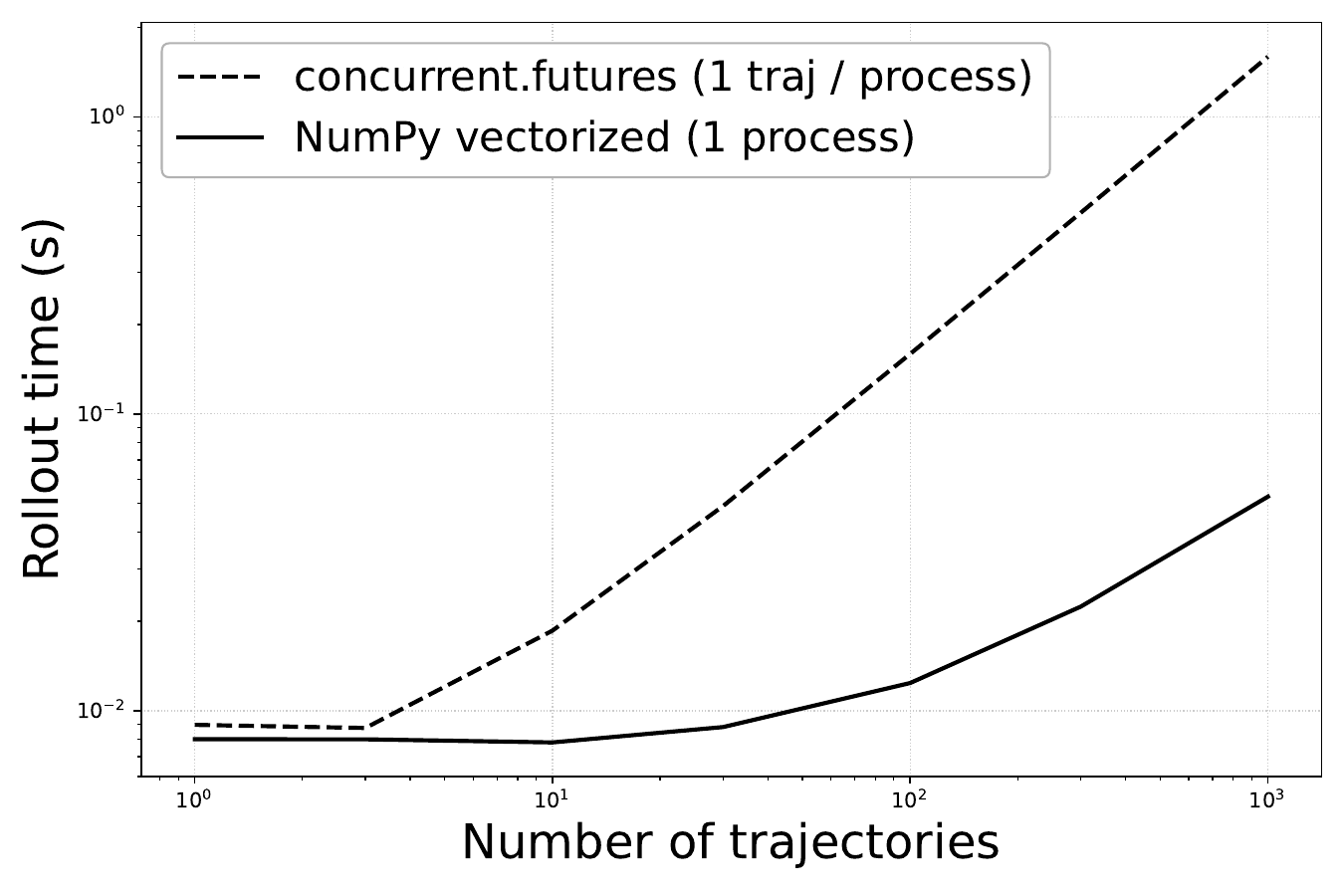}
    \caption{Speedup of vectorization using NumPy.}
    \label{fig:speedup}
\end{figure}

\subsection{Vectorized approach}
\label{subsec:vectorized-approach}

RL agents require training policies across many market paths in order to obtain reliable gradient estimates and reduce sensitivity to any single price or order-flow realization. To support this at scale, our implementation adopts a vectorized simulation architecture in which all state variables, actions, rewards, and the associated stochastic processes presented in \Cref{sec:design-module} carry a leading batch dimension of size \texttt{num\_trajectories}, the number of rollouts simulated in parallel, using NumPy arrays \cite{van_der_Walt_2011}. Consequently, each environment step produces \texttt{num\_trajectories} transitions.\footnote{This vectorization is native to the AMM structure, and it differs from Gymnasium generic \texttt{VectorEnv} , which is agnostic to the AMM simulation engine.}
Our design contrasts with the typical approach to parallelize RL ``rollouts'' at the process level, which instantiates many environments across multiple CPU cores and sequentially generates trajectories on each one. Our implementation, instead, amortizes memory allocation and mathematical operations over the full batch in a single pass. 
\Cref{fig:speedup} illustrates the speedup offered by our NumPy implementation over the multiprocessing approach that uses the concurrent.futures package for identical per-trajectory workload.\footnote{The benchmark was run on a single machine with 8 CPU cores and 16 GB of RAM.} For rolling out 1000 trajectories, the vectorized approach takes $\approx 0.05$ s versus $\approx 1.6$ s for the multiprocessing approach, achieving a speedup of roughly 30×. Typical RL training involves millions of rollouts, making the multiprocessing approach prohibitively slow. 


\subsection{Reinforcement Learning under Domain Randomization}
\label{subsec:domain-rnd}
To demonstrate the potential of \saife , we consider the evaluation of liquidity provision policies under distributional shifts in market parameters. As a warmup, we provide an example of a nominal RL agent, under \textit{stationary transition dynamics}, whereby the learning agent is trained and evaluated under the same market parameters. 
Subsequently, we proceed to domain randomization, whereby we sample market parameters during training, a process that we call \textit{non-stationary transition dynamics}.  

\paragraph{Stationary Transition Dynamics}
As a first example of training an agent using \saife , we use proximal policy optimization (PPO) \cite{Schulman17ppo} to train an LP agent in a \clmm under fixed market parameters. 
In this example, the external midprice $S_t$ follows a geometric Brownian motion \eqref{eq:GBM}, with $\mu=0$ and $\sigma=0.03$. The evolution of the marginal exchange rate $Z_t$ is formed by incoming LT trades, for which we use the intensity model in \eqref{eq:linear_intensities}, with $\alpha_1=450, \alpha_2 = 10$ and $\alpha_3=4000$, where the latter captures the arbitrage pressure and is the main driver through which the marginal exchange rate $Z_t$ follows the external midprice $S_t$ throughout the simulation period. We use the depth-dependent price impact model described in \Cref{subsec:price-impact}. 
We set the initial price at $Z_0=S_0=100$ and the LP agent's initial wealth at $1000$ in asset X. We initialize the liquidity depth profile \eqref{eq:total-liquidity} with $100,000$ liquidity units in each tick range. Near the tick ranges of the initial marginal exchange rate at $100$, these liquidity units correspond to roughly $50$ units in asset $X$. Moreover, we fix the size $Q$ of each LT trade at $250$ units in asset $X$, which causes a price impact of around 5 ticks in the marginal exchange rate $Z$. We set a fixed gas cost at $2$ units in asset $X$.

\begin{figure}[t]
    \centering
    \includegraphics[width=\linewidth]{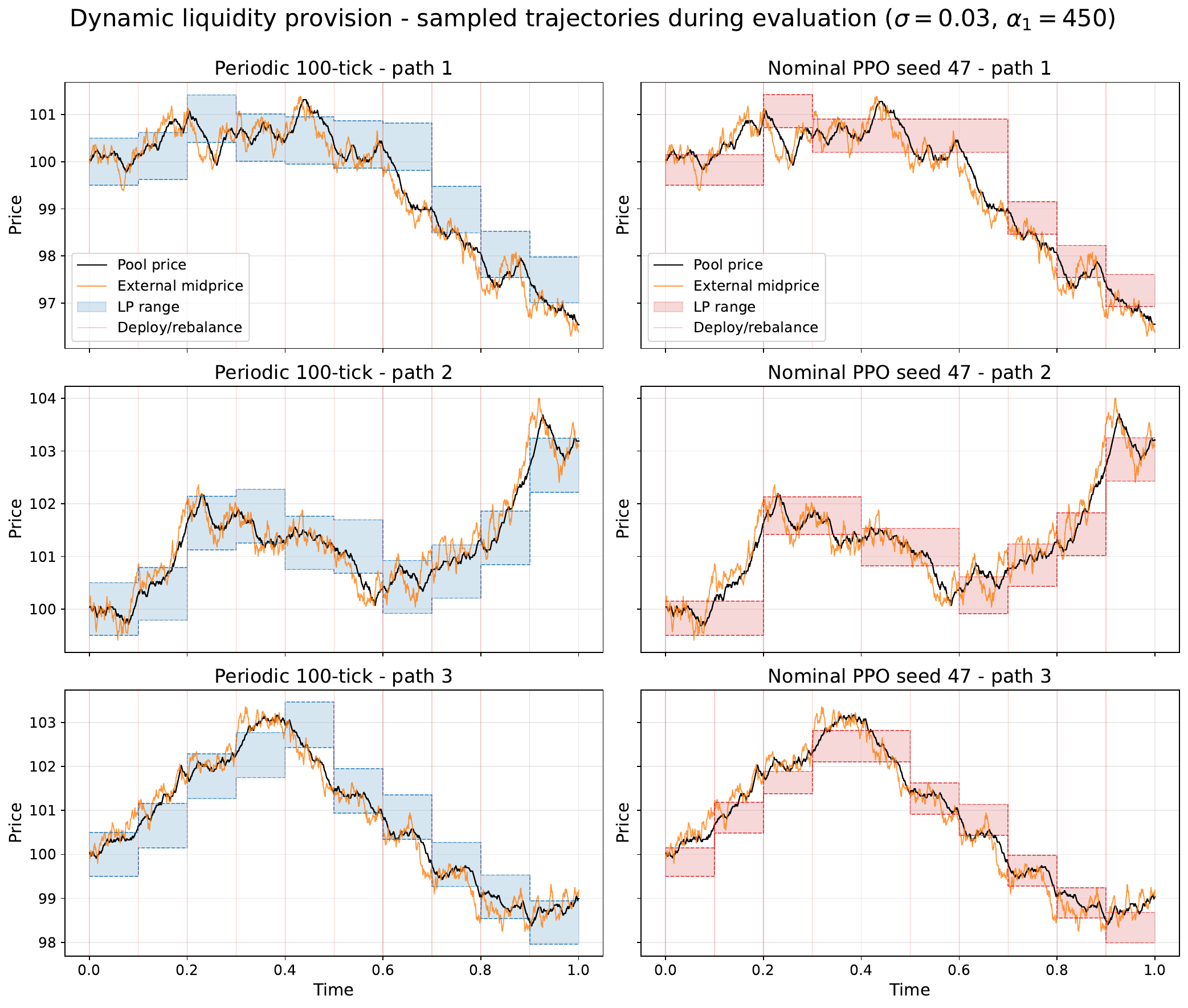}
    \caption{Dynamic liquidity provision policies.  On the left column, we plot the strategy of the baseline agent (\textit{periodic rebalancing}), that rebalances symmetrically around the marginal exchange rate. On the right column, we plot sampled trajectories from the learnt policy of the PPO agent. This is for illustrative purposes to visualize how the agent reacts to market conditions.}
    \label{fig:sampled-trajectory}
\end{figure}

We allow the agent to take one action every $100$ time steps through the wrapper \texttt{DecisionStrideEnv}, and we allow the agent to choose her new positions within an interval of $50$ tick ranges from the current marginal exchange rate. The restrictions on the agent's action space are natural, but, importantly, they improve the agent's exploration. As a reward class, we choose the running inventory penalty from \eqref{eq:running_inv}, setting the per-step inventory penalty at $\phi=0.4$, the terminal inventory penalty at $\eta=0$, and the target level at $\hat{y}=0$. We compare the performance of the RL agent with that of a baseline agent that rebalances every $100$ time steps symmetrically around the current marginal exchange rate $Z$, with $50$ tick ranges on each side. 
We spawn $1000$ trajectories that run in parallel, and we sample three of those during evaluation, as illustrated in \Cref{fig:sampled-trajectory}.

\paragraph{Non-Stationary Transition Dynamics}
In order to expose the agent to market uncertainty, we implement episode-level domain randomization through a wrapper around the base environment class \texttt{AMMEnvironment}. We consider the market domain $\theta=(\sigma,\alpha_1)$, where $\sigma$ is the volatility of the external midprice process $S_t$, and $\alpha_1$ is the baseline arrival intensity in \eqref{eq:linear_intensities}, which captures the intensity of noise trading. During training, these market parameters are sampled from the joint distribution $\rho_{\theta}$ for the independent random variables $\sigma, \alpha_1$, according to $\sigma\sim\mathcal{U}(0.01,0.05)$ and $\alpha_1\sim\mathcal{U}(300,600)$.

At the start of each rollout, the wrapper samples $D$ domains; that is, $D$ tuples of $(\sigma, \alpha_1)$ and assigns them into balanced groups across the $n$ trajectories that run in parallel. 
The wrapper creates $D$ groups, each containing $n/D$ trajectories, and uses the same domain within the group throughout the episode. The hyperparameter $D$ offers the following simple tradeoff. Larger $D$ gives each PPO update broader coverage of the domain distribution. Smaller $D$ gives more repeated stochastic trajectories per sampled domain. The remaining market parameters are kept fixed across domains. 

Naturally, the market domain $\theta = (\sigma, \alpha_1)$ is not included in the RL agent's observation space; thus, the agent faces a \textit{latent-parameter} partially observable decision problem.
Consequently, the domain randomized agent must learn behaviors that perform well across the domain distribution. 
We use the same market configuration as in the \textit{stationary transition dynamics} example in the previous paragraph.
The reward is the running-inventory criterion defined in \eqref{eq:running_inv} with a per-step inventory penalty $\phi=0.4$. The objective is modified to maximize the expected reward across the distribution $\rho_\theta$ 
\begin{equation}
\label{eq:domain-randomize-criterion}
\mathbb{E}_{\theta \sim \rho_{\theta}} \left[
\mathbb{E}_{\traj \sim p(\traj |\pi, \theta) }
\left[
\sum_{n=0}^{N-1}
\left(
V_{n+1}-V_n
-\phi\,\Delta t\,y_{n+1}^{2}
\right)
\right]
\right],
\end{equation}
where the trajectory $\traj$ is sampled from the trajectory distribution induced by policy $\pi$ in the environment with the market domain $\theta$.

\begin{figure*}
    \centering
    \includegraphics[width=\linewidth]{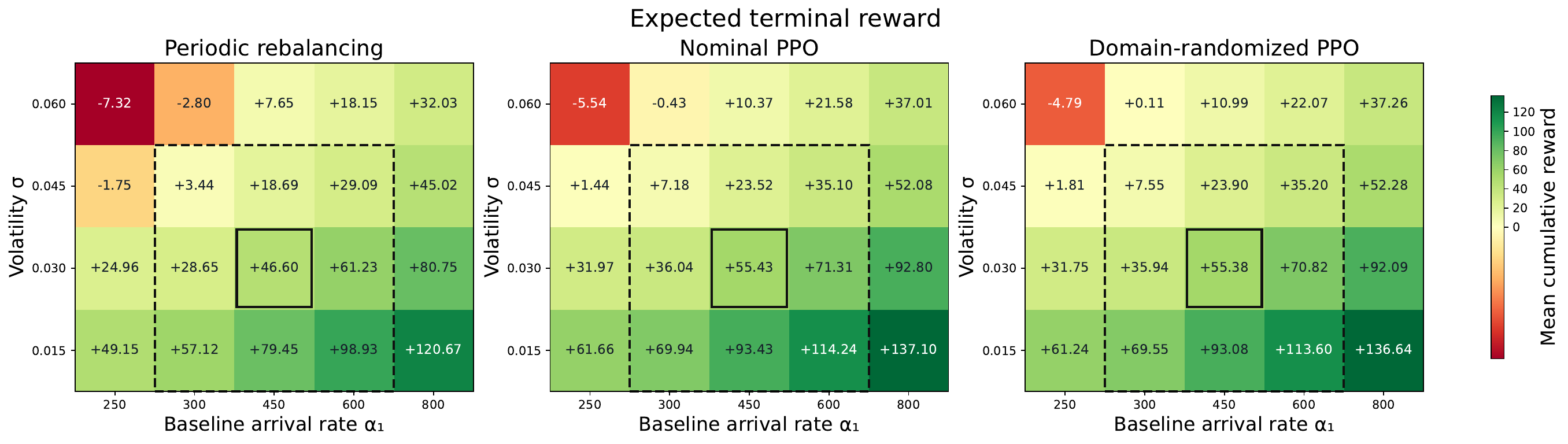}
    \caption{Evaluation of the agents under different market regimes. The left panel shows the periodic-rebalancing baseline agent, the middle panel shows the nominal PPO agent, trained in the market domain $(\sigma,\alpha_1)=(0.03,450)$ and the right panel shows the domain-randomized agent, trained on market domains sampled from the distribution $(\rho_{\theta})$. The evaluation covers both in-distribution market parameters (the dashed area) and out-of-distribution market parameters.}
    \label{fig:heatmaps-nominal-random-periodic}

\end{figure*}

For evaluation, we compare the domain-randomized agent with the \textit{nominal} PPO, trained under the fixed domain $(\sigma,\alpha_1)=(0.03,450)$. Policies are evaluated using Monte Carlo trajectories at each of the 20 parameter points $\{0.015, 0.03, 0.045,0.06\}\times\{250,300,450, 600, 800\}$, which include in-distribution parameters as well as a few out of distribution market parameters for stress testing. We report the mean objective within each regime, which is illustrated in \Cref{fig:heatmaps-nominal-random-periodic}. We note that the domain-randomized agent uses a feedforward neural-network policy. An alternative approach would be to implement a recurrent policy that learns to infer dynamics from observation and action history, as is done in \cite{Peng18sim}.


\section*{Conclusion and Further Research}
In this paper, we introduce \saife , a collection of RL environments for solving trading problems in constant product markets with concentrated liquidity. The goal was to assemble the components of the AMM microstructure in a modular way, whereby researchers and practitioners can simulate their desired environment.  We demonstrate the versatility of \saife by training and evaluating RL agents under different market parameters. The vectorized design of the environments makes these experiments computationally tractable and allows policies to be
evaluated across many trajectories under controlled variations of the underlying market model. 

While \saife provides a solid foundation for research, there are several  natural directions for future work. A first extension is to develop multi-agent RL environments. In practice, liquidity provision is strategic, and, unlike market making in limit order books, it is highly transparent. Hence, multiple LPs compete for order flow by responding to common market signals and other LPs' actions \cite{baggiani2026competition}. A second direction is to further expand the action space of LPs in our gym environment. In particular, our implementation is forward-compatible, allowing LPs to set dynamic trading fees \cite{baggiani2025optimal}. 
Last but not least, we invite the community to implement additional agents from the LT perspective. Our \texttt{SpeedControlArbitrageurAgent} is a first building block in this direction. Such extensions would make it possible to study settings that interpolate between the model-based environments developed in this work and more agent-based simulations, in which market dynamics emerge from the interaction of heterogeneous LPs, arbitrageurs, and noise
traders rather than from exogenously specified stochastic processes. 

\begin{acks}
This work was supported by UK Research and Innovation (UKRI) through the Engineering and Physical Sciences Research Council (EPSRC) under grant EP/W034042/1 (SAiFE: A Sandbox for AI-powered Trading in Decentralised Financial Markets). Carmine Ventre was partially supported by UKRI (Grant number UKRI4151).  
\end{acks}

\bibliographystyle{ACM-Reference-Format}
\bibliography{bibliography}

\end{document}

%% file: defs.tex
\newcommand{\defeq}{\vcentcolon=}  
\newcommand{\reals}{{\mbox{\bf R}}}
\newcommand{\ie}{{\it i.e.}}
\newcommand{\Rn}{\mathcal{R}_{+}^n}
\newcommand{\D}{\Delta}

\newcommand{\la}{\lambda}
\newcommand{\de}{\delta}
\newcommand{\ka}{\kappa}

\newcommand{\al}{\alpha}

\newcommand{\tka}{\tilde{\kappa}}
\newcommand{\tKa}{\tilde{\Kappa}}

\newcommand{\CPM}{Constant Product Market}

\newcommand{\IL}{\text{IL}}

\newcommand{\fee}{\tau}
\newcommand{\pen}{\phi}
\newcommand{\win}{\rho}
\newcommand{\imp}{r}
\newcommand{\traj}{\zeta}

\newcommand{\E}{\mathbb{E}}
\newcommand{\dd}{\,d}

\newcommand{\saife}{\texttt{SAiFE\_gym}\xspace}

\newcommand{\counts}{\mathcal{N}}

\NewDocumentCommand{\z}{e{^}}{%
  Z\IfValueT{#1}{^{(#1)}}%
}

\newcommand{\clmm}{CPM with CL\xspace}
\newcommand{\clmms}{CPMs with CL\xspace}